\documentclass[twocolumn,preprintnumbers]{revtex4}
\usepackage{amsmath}
\usepackage{amssymb}
\usepackage{graphicx}
\usepackage{bm}

\begin{document}


\title{A Statistical Model of Magnetic Islands in a Large Current Layer}

\author{R.~L.~Fermo}
\author{J.~F.~Drake}
\author{M.~Swisdak}
\affiliation{IREAP, University of Maryland, College Park, MD 20742-3511, USA}

\date{\today}

\begin{abstract}

We develop a statistical model describing the dynamics of magnetic
islands in very large current layers that develop in space plasma.
Two parameters characterize the island distribution: the flux $\psi$
contained in the island and the area $A$ it encloses.  We derive an
integro-differential evolution equation for this distribution
function, based on rules that govern the small-scale generation of
secondary islands, the rates of island growth, and island merging.
Our numerical solutions of this equation produce island distributions
relevant to the magnetosphere and corona.  We also derive and
analytically solve a differential equation for large islands that
explicitly shows the role merging plays in island growth.

\end{abstract}

\maketitle

Recent reconnection simulations have shown that guide field
reconnection becomes bursty \cite{Drake06}.  Electron current layers
near the magnetic x-line lengthen and become unstable to the formation
of secondary magnetic islands, which dynamically grow in size and
begin to merge.  The existence of these magnetic islands is
consistent with observations in both the magnetotail and the
magnetopause \cite{Russell79}, and their occurrence in current layers
on the sun has also been inferred \cite{Sheeley04}.  The presence of
many islands could also account for the observed thickness of current
sheets associated with coronal mass ejections (CMEs), which is far
greater than that predicted by reconnection models employing classical
or anomalous resistivity \cite{Ciaravella08}.

Magnetic island formation is of great interest because of its impact
on reconnection rates and because both simulations \cite{Drake06a} and
observations \cite{Chen08} indicate that islands are correlated with
highly energetic electrons.  However, the dynamics of islands in very
large-scale current layers are not yet well-understood.  On the one
hand, even the largest particle-in-cell (PIC) simulations
\cite{Daughton09} come nowhere near the scale sizes of these current
layers ($L \sim 4000 \,d_i$ in the magnetopause, $L \sim 10^6 \,d_i$
in the corona).  On the other hand, fluid simulations that encompass
these large scales\cite{Raeder06} fail to capture the small-scale
dynamics of reconnection within the dissipation region.

In light of these limitations, we employ a novel statistical method
for treating two-dimensional magnetic islands in large current layers.
We define a statistical distrubtion function which describes islands
in the whole current layer and develop an evolution equation for that
distribution.  We then present steady-state solutions that show how
the merger of smaller islands drives the growth of large islands.  The
model is based on rules of island merging based on two-dimensional
particle-in-cell reconnection simulations.

We model the islands using a distribution function $f$ in terms of the
island's flux $\psi$ and area $A$.  Defined in this manner, $dN =
f(\psi,A)d\psi dA$ signifies the number of magnetic islands with flux
in the range $[\psi,\psi+d\psi]$ and area in $[A,A+dA]$.  The
distribution function $f(\psi,A,t)$ evolves in time in the phase space
of $(\psi , A)$, independent of a space coordinate, because it
describes islands over the whole current sheet of length $L$.  The two
quantities $\psi$ and $A$ are sufficient to completely characterize
the state of an island: the magnetic field strength is computed as $B
= \psi/r = \psi\sqrt{\pi/A}$, where $r = \sqrt{A/\pi}$ is a
characteristic radius.

With the distribution function defined, we now establish rules for the
behavior of magnetic islands, and formulate how these rules affect the
distribution.  Islands first form at scales between the electron $d_e$
and ion $d_i$ skin depths in the current layers near x-lines due to the
tearing instability \cite{Drake06}.  The evolution equation therefore
includes a source term $S(\psi , A)$ at this scale, in our case of
Gaussian shape.  As islands convect outwards at the Alfv\'en speed, a
sink term models the convection of islands out of the system at the
rate $c_A/L$.

Furthermore, the model must account for the growth of magnetic
islands due to reconnection.  PIC simulations have shown that once
reconnection reaches a nonlinear stage, the normalized reconnection 
rate plateaus at roughly $\varepsilon \approx 0.1$ \cite{Shay07}.
Given this constant reconnection rate and an asymptotic reconnection
magnetic field of $B_0$, the magnetic flux of an island increases at 
the rate $\dot{\psi} = v_{in}B_x = \varepsilon c_A B_0$.  Likewise, the
island's characteristic radius $r$ increases at a constant rate
$\dot{r} = \varepsilon c_A$, and so the area increases at the rate
$\dot{A} = 2\pi rv_{in} = 2\varepsilon c_A \sqrt{\pi A}$.

\begin{figure}
\noindent\includegraphics[width=3.0in]{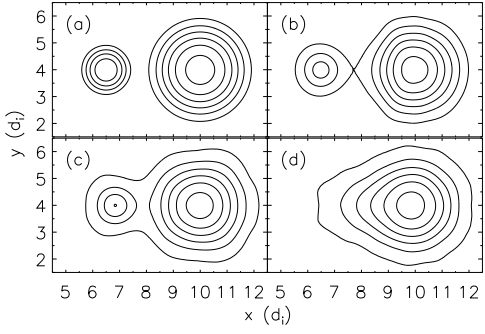}
\caption{\label{merging} The results of a PIC simulation of islands
merger, showing the magnetic field lines at (a)
$t=0\,\Omega_{ci}^{-1}$, (b) $t=0.3\,\Omega_{ci}^{-1}$, (c)
$t=4.0\,\Omega_{ci}^{-1}$, and (d) $t=8.0\,\Omega_{ci}^{-1}$, where
$\Omega_{ci}$ is the ion cyclotron frequency. The smaller island
initially has 75\% of the flux and 25\% of the area of the larger
island. }
\end{figure}

We now describe the rules for island merger: the merger of two islands
yields an island with an area $A$ equal to the sum of the two initial
areas and a flux $\psi$ equal to the higher of the two initial fluxes.
The flux does not add because magnetic reconnection does not increase
or decrease magnetic flux but simply changes its connectivity.  Thus,
field lines from the island with more flux reconnect with those from
the island with less flux until all of the latter's flux is depleted.
The simplicity of these rules reflect our choice of $\psi$ and $A$ as
the variables defining our phase space.  Fig.\ \ref{merging} shows
results from a PIC simulation that demonstrates this in the simple
case of two isolated flux bundles.  Magnetic field contours are shown
at various times during the merging process, with the outermost field
line representing the boundary of the island for the purposes of
computing area.  Both the maximum flux $\psi$ and the total area $A$
remained nearly constant throughout the simulation, with variations
having standard deviations of $2\%$ and $7\%$ of the mean,
respectively.  These merging rules reveal why the merging process is
energetically favorable: the dissipation of magnetic energy in the
reconnection process.  We can write the magnetic energy in an island
as $W = B^2A \sim \psi^2$.  Before the merger the energy is given by
$W_i \sim \psi_1^2 + \psi_2^2$, and after the merger the energy
(supposing $\psi_1 \le \psi_2$ without loss of generality) is $W_f
\sim \psi_2^2 < W_i$.

We now develop an equation for the island distribution $f$ that
parallels the derivation of the collisional Boltzmann equation. In the
absence of island merging the number density of magnetic islands is
preserved as reconnection changes an island's area and flux, while
merging (analgous to collisional scattering in the case of the
Boltzmann equation) causes a jump in the local island number. We first
write the merging velocity of two islands with fluxes $\psi_1, \psi_2$
and areas $A_1, A_2$, assuming constant mass density $\rho$,
\begin{equation}
v^2(\psi_1,A_1,\psi_2,A_2) =
\dfrac{\varepsilon^2}{4\pi\rho}\dfrac{\psi_1\psi_2
r_1r_2(r_1^2+d_i^2)^{1/2}(r_2^2+d_i^2)^{1/2}}
{(r_1^2+d_e^2)^{3/2}(r_2^2+d_e^2)^{3/2}}
\end{equation}
where the dependence on $A_1$ and $A_2$ is implicit in $r_1$ and
$r_2$.  For islands larger than $d_i$, this formula resembles the
hybrid Alfv\'en velocity for asymmetric reconnection, $v^2 = B_1 B_2 /
4 \pi \rho$ \cite{Cassak07c,Swisdak07}.  For smaller islands, down to
the electron skin depth $d_e$, $v^2 \approx \psi_1 \psi_2 d_i^2 / 4
\pi \rho r_1^2 r_2^2$ in agreement with the dispersion relation
$v_{ph} \propto k$ of the whistler dynamics that dominate this regime.
Lastly, $v \rightarrow 0$ as $r \rightarrow 0$.

Now we consider the number of islands $\Delta N\big{|}_{\mbox{src}}
(\psi , A)$ formed by merging. A merged island with flux $\psi$ arises
from an island of flux $\psi$ that has merged with another island with
flux $\psi' \le \psi$.  Likewise, an island with area $A$ must come
from an island with $A' < A$ that has merged with an island of area $A
- A'$.  The probability of those two islands merging within a time
$\Delta t$ is given by $v(\psi,A',\psi',A-A')\Delta t/L$, so
\begin{equation}
\Delta N\big{|}_{\mbox{src}} = \int_0^AdA'f(\psi,A')
\int_0^{\psi}d\psi'\dfrac{v\Delta t}{L}f(\psi',A-A')\Delta\psi\Delta
A\,.
\end{equation}

A similar analysis yields the number of islands lost through merging
$\Delta N\big{|}_{\mbox{snk}}(\psi , A)$.  Such an island is lost if
it merges with any island of finite flux $\psi'$ and area $A'$.
Again, the probability of merger depends on $v(\psi,A,\psi',A')\Delta t/L$,
and
\begin{equation}
\Delta N\big{|}_{\mbox{snk}} = -\int_0^\infty dA'\int_0^\infty
d\psi'\dfrac{v\Delta t}{L}f(\psi',A')f(\psi,A)\Delta\psi\Delta A\,.
\end{equation}

The rate of change of $f$ due to merging is then calculated as
$(\Delta N\big{|}_{\mbox{src}}+\Delta
N\big{|}_{\mbox{src}})/\Delta\psi\Delta A\Delta t$. Combining this
expression with the change in $f$ due to reconnection, we get our
evolution equation:
\begin{align}\label{evolution}
\lefteqn{\frac{\partial f}{\partial t} + 
\frac{\partial}{\partial \psi}(\dot{\psi}f) + 
\frac{\partial}{\partial A}(\dot{A}f) 
= S(\psi,A) - \frac{c_A}{L}f } \nonumber\\
& {} + \frac{1}{L}\int_0^AdA'f(\psi,A')\int_0^\psi
d\psi'v(\psi,A',\psi',A-A')f(\psi',A-A') \nonumber\\ 
& {} - \frac{1}{L}f(\psi,A)\int_0^\infty dA'\int_0^\infty
d\psi'v(\psi,A,\psi',A')f(\psi',A')\,.
\end{align}
On the left-hand side, the time-derivative of $f$ results from the
change in flux $\dot{\psi}$ and area $\dot{A}$ due to reconnection,
which act to increase the island size and therefore conserve total
island number.  On the right-hand side, we have a source $S(\psi,A)$,
a sink representing island convection out of the system, and the
merging terms. Consistent with the merging rules, the merging terms in
Eq.\ (\ref{evolution}) preserve total area, which can be shown by
multiplying by $A$ and integrating over $A$ and $\psi$.

\begin{figure}
\noindent\includegraphics[width=3.0in]{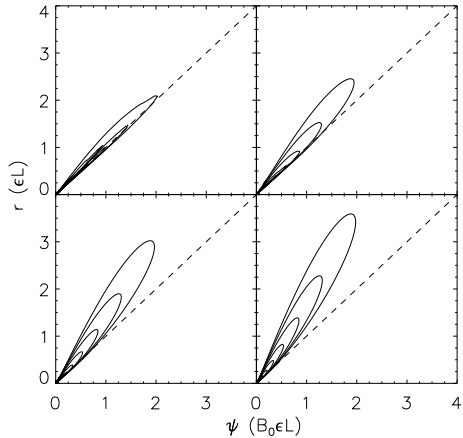}
\caption{\label{varySn} The steady-state distribution function
$F_\infty(\psi,r)$ for (a) $S_N^{\,~*} = 4$, (b) $S_N^{\,~*} = 40$,
(c) $S_N^{\,~*} = 400$, and (d) $S_N^{\,~*} = 4000$.  The contours are 
logarithmic. }
\end{figure}

We first consider the case where the merging terms are negligible.
This is valid if there are few islands in the layer ($S(\psi,A)$ is
small).  For simplicity, we change variables from the area $A$ to a
characteristic radius $r$, with $A = \pi r^2$ (although the islands
need not be circular) and the new distribution function,
\begin{equation}
F(\psi,r) = f(\psi,A)\dfrac{dA}{dr} = 2\pi rf(\psi,\pi r^2)\,.
\end{equation}
The evolution equation, without merging, becomes
\begin{equation}\label{nomerge}
\dfrac{\partial F}{\partial t} + 
\dfrac{\partial}{\partial\psi}(\dot{\psi}F) + 
\dfrac{\partial}{\partial r}(\dot{r}F) 
= S(\psi,r) - \dfrac{c_A}{L}F \,.
\end{equation}
With a delta function source $S=\delta(\psi-\psi')\delta(r-r')$, the
steady-state solution $F_G$ is given by the Green's function
\begin{align}\label{greensfn}
F_G(\psi,r,\psi',r') = & \dfrac{1}{\varepsilon c_A}
\exp{\left(-\dfrac{r-r'}{\varepsilon L}\right)} H(r-r') \nonumber\\
& {} \times \delta\left((\psi-\psi')-B_0(r-r')\right)
\end{align}
where $H(r)$ is the Heaviside function.  All islands in this solution
have a magnetic field strength $B_0$, equal to the ambient magnetic
field.  The solution in $\psi-r$ phase space is a decaying exponential
along the line $\psi=B_0r$, starting at $\psi=\psi'$ and $r=r'$. The
corresponding solutions for $f$ are shown in dashed lines in
Fig.~\ref{varySn}.  A key feature of Eq.\ (\ref{greensfn}) is the
characteristic island size $r_0 = \varepsilon L$, which arises from
balancing the island growth rate of $\dot{r} = \varepsilon c_A$ with a
system transit time of $L/c_A$.  For example, with a reconnection rate
of $\varepsilon \sim 0.1$ on the magnetopause where $L \sim 30\,R_E$,
this simple model predicts islands of size $\sim 3\,R_E$.  A survey of
flux transfer events along the magnetopause \cite{Rijnbeek84,
Saunders84} determined that typical scale sizes are $0.5\,R_E \times
2\,R_E$, for which $r \approx 1\,R_E$.  Such island sizes are also
consistent with those seen in direct observations of current sheets
formed by CMEs \cite{Lin05}.

If the merging terms are kept, the resulting integro-differential
equation must be solved computationally.  Although Eq.\
(\ref{evolution}) appears to contain several parameters (the current
sheet length $L$, the reconnection rate $\varepsilon$, and the source
term strength $S_N = \int_0^\infty dr \int_0^\infty d\psi S(\psi,r)$
which gives the rate of island nucleation), in reality it only has one
free parameter.  By normalizing time to $L/c_A$, island flux to
$\varepsilon B_0 L$, island size to $\varepsilon L$, and the merging
velocity to $\varepsilon c_A$, the only free parameter is the
normalized amplitude of the source, $S_N^{\,~*} = S_N\varepsilon
L/c_A$.  Although we set $L = 4000\,d_i$ here (a realistic value for
the magnetopause), we can essentially model any system by varying
$S_N$.

The integration of Eq.~(\ref{evolution}) yields the steady state
island distribution, $F_{\infty}(\psi,r)$, which is shown for four
values of $S_N$ in Fig.\ \ref{varySn}.  Whereas the solution without
merging remains localized along the $\psi=B_0r$ diagonal, the merging
terms break this symmetry: the merging process increases area but not
flux so the distribution functions curve away from the diagonal toward
larger $r$.  One test of the veracity of our numerical solutions is
their consistency with the moments of the evolution equation.
Integrating Eq.\ (\ref{evolution}) in $\psi$ and $A$ yields an
equation for the total number of islands $N = \int_0^\infty dA
\int_0^\infty d\psi f(\psi,A,t)$:
\begin{equation}\label{mergeNeqn}
\dfrac{dN}{dt} = S_N - \dfrac{c_A}{L}N - \dfrac{\varepsilon c_A}{2L}N^2
\end{equation}
where, for simplicity, we have taken the merging velocity to be
$\varepsilon c_A$.  The merging terms reduce $N$ since two islands
merge into one and are responsible for the term quadratic in $N$.
Eq.\ (\ref{mergeNeqn}) can be solved analytically:
\begin{equation}\label{mergeN}
N(t) = \dfrac{N_f \left(1 - e^{-t/t_s}\right)}
{1 + (\varepsilon^2 N_f^2/2S_N^{\,~*}) e^{-t/t_s}}
\end{equation}
where
\begin{equation}\label{Nfinal}
N_f = \dfrac{1}{\varepsilon} 
\left[\left(1 + 2S_N^{\,~*}\right)^{\frac{1}{2}} -1 \right],
t_s = \dfrac{L}{c_A} \left(1 + 2S_N^{\,~*}\right)^{-\frac{1}{2}}
\end{equation}
are the asymptotic number of islands in steady state and a time-scale
to reach it.  The form of $N(t)$ predicted by Eq.\ (\ref{mergeN})
compares favorably to that of the numerical solution.  Note from Eq.\
(\ref{Nfinal}) that $S_N$ dictates the number of islands in the
system, and so it effectively controls the importance of merging.  For
larger $S_N$, the merging term ($\sim N^2$) in Eq.~(\ref{mergeNeqn})
dwarfs the convective loss term ($\sim N$).

Similarly, by multiplying Eq.\ (\ref{evolution}) by $A$ and
integrating, we get a moment equation for the total area of all the
islands in the system, $A_T = \int_0^\infty dA \int_0^\infty d\psi A
f(\psi,A,t)$:
\begin{equation}\label{Atot}
\dfrac{dA_T}{dt} = 2\pi\varepsilon c_A r_T - \dfrac{c_A}{L}A_T
\end{equation}
where $r_T = \int_0^\infty dr \int_0^\infty d\psi rF(\psi,r,t)$ and
the (small) source contribution was neglected.  Since merging
conserves total area, the merging integrals do not contribute to this
equation.

\begin{figure}
\noindent\includegraphics[width=3.0in]{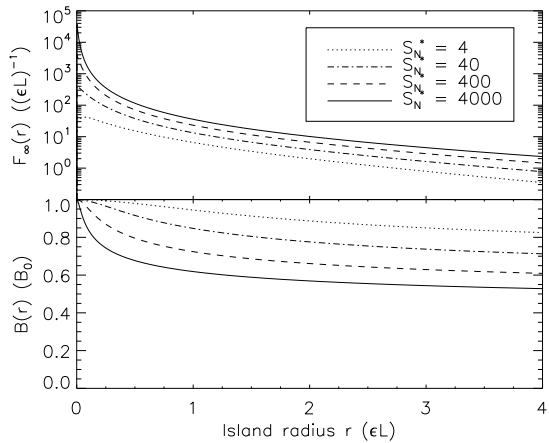}
\caption{\label{distrAvgB} (a) The steady-state distribution function
in $r$, given by $F_\infty(r)$ as defined in Eq.\ (\ref{fr}) for
various $S_N$. (b) The average magnetic field strength $B$ as a
function of island radius $r$ for various $S_N$. }
\end{figure}


Fig.\ \ref{distrAvgB}(a)
shows the distribution of islands in radius,
\begin{equation}\label{fr}
\bar{F}_\infty(r) = \int_0^\infty d\psi F_\infty(\psi,r)
\end{equation}
for each $S_N$.  The behavior of these curves for the largest islands
can be deduced from Eq.\ (\ref{evolution}) for large $A$.  Integrating
over $\psi$, assuming a constant merging velocity $v =\varepsilon
c_A$, and expanding the integral over the merging terms gives an
equation for $\bar{f}(A,t) = \int_0^{\infty}d\psi f(\psi,A,t)$:
\begin{equation}\label{larger}
\frac{\partial \bar{f}}{\partial t} + \frac{\partial}{\partial
A}(\dot{A}\bar{f}) + \frac{\varepsilon c_A
A_T}{L}\frac{\partial \bar{f}}{\partial A}= -\frac{c_A}{L}\bar{f} 
\end{equation}
The third term on the left side, which arises from the merging terms,
describes how large islands grow in area by devouring smaller islands
-- the coefficient $\varepsilon c_A A_T/L$ is the rate at which the
total area $A_T$ of all the smaller islands is consumed.  By balancing
the second and third terms on the left side we obtain a characteristic
length scale above which growth via reconnection dominates growth via
merging.  Recalling that $\dot{A} = 2 \pi \varepsilon c_A r$, we find
that the transition occurs at $\hat{r} = A_T/2\pi L$. In steady state
Eq.~(\ref{Atot}) yields $A_T = 2\pi \varepsilon L r_T$ so $\hat{r} =
\varepsilon r_T$.

Equation (\ref{larger}) admits a steady state solution,
\begin{equation}\label{largesol}
\bar{F}_{\infty}(r) = C e^{-r/\varepsilon L} \, r (r + \varepsilon r_T)^{r_T/L - 1},
\end{equation}
where $C$ is an arbitrary constant. For sufficiently large $r$, the
exponential behavior of Eq.\ (\ref{largesol}) will dominate.  In all
four cases shown in Fig.\ \ref{distrAvgB} the distribution of islands
at large radii agrees with the expression given in Eq.\
(\ref{largesol}).

Merging increases an island's area but not its flux, and hence leads
to a decreased in-plane field strength $B = \psi/r$.  Since merging
scales as $N^2$, one might surmise that the more islands in the system
(e.g., because of larger $S_N$), the more merging takes place, and the
smaller the magnetic field $B$.  This hypothesis is borne out in Fig.\
\ref{varySn}, in that the steady-state solution for larger $S_N$ tilts
farther away from the $\psi = B_0 r$ diagonal.  This effect is seen in
Fig.\ \ref{distrAvgB}(b), which shows, for various $S_N$, the average
magnetic field strength as a function of island radius
\begin{equation}\label{br}
B(r) = \dfrac{\int_0^\infty d\psi \dfrac{\psi}{r} F_\infty(\psi,r)}
{\int_0^\infty d\psi F_\infty(\psi,r)} \,.
\end{equation}

Comparisons of the present results with those of recent MHD
simulations of large current layers \cite{Samtaney09} are not possible
since these simulations were limited to the early time behavior, where
island merging plays a minimal role.  Therefore, in future work we
will benchmark this model using Hall MHD simulations to justify our
equations for a current layer with many islands.  Although
observational data (for example, from THEMIS at the magnetopause)
cannot directly measure $S_N$, the known distributions of island sizes
and magnetic field strengths could, using our model, be used to infer
it. An encouraging early result is a histogram of 1223 FTEs detected
by THEMIS between 2007 and 2008, which shows an exponential decrease
in event number for island scale lengths above an $R_E$
\cite{Zhang09}. Similarly, TRACE observations of supra-arcade
downflows in the solar corona during flares could also yield
distributions of island size \cite{McKenzie09}.

\smallskip

This work was supported by NASA through an Earth and Space Science
Fellowship NNX07A083H and NNX09A102G.  Computations were performed at
the National Energy Research Scientific Computing Center.


%

\end{document}